\documentclass[graybox, envcountchap]{svmult}

\usepackage{mathptmx}        % selects Times Roman as basic font
\usepackage{amsmath}
\usepackage{amssymb}
\usepackage{mathrsfs} 
\usepackage{bm} 
\usepackage{color}
\usepackage{helvet}          % selects Helvetica as sans-serif font
\usepackage{courier}         % selects Courier as typewriter font
\usepackage{dirtree}
\usepackage{makeidx}        % allows index generation
\usepackage{graphicx}        % standard LaTeX graphics tool
\usepackage{subfig}

\usepackage{multicol}        % used for the two-column index
\usepackage[bottom]{footmisc}% places footnotes at page bottom

\usepackage{hyperref}        %for hyperlinks
\hypersetup{colorlinks=true,urlcolor=blue, citecolor    = gray}
\usepackage{xcolor}
\usepackage{doi}
\usepackage{url}
\usepackage[misc]{ifsym}
\usepackage{ulem}

\makeindex             % used for the subject index
\usepackage{comment}
\usepackage{mathrsfs}
\begin{document}
\title{Singularity-free gravitational collapse: From regular black holes to horizonless objects\vspace*{-3cm}\\
{\hspace*{10cm}{
\small{YITP-23-05}}}\vspace*{3cm}}
%Chapter Title
\titlerunning{{Singularity-free gravitational collapse: From RBHs to horizonless objects}}
%for an abbreviated version of
\author{Ra\'ul Carballo-Rubio, Francesco Di Filippo, Stefano Liberati,\\ and Matt Visser}
%\authorrunning{Carballo-Rubio, Di Filippo, Liberati, Visser}
\institute{Ra\'ul Carballo-Rubio (\Letter) \at CP3-Origins, University of Southern Denmark, Campusvej 55, DK-5230 Odense M, Denmark and Florida Space Institute, University of Central Florida, 12354 Research Parkway, Partnership 1, 32826 Orlando, FL, USA, \email{raul@sdu.dk}
\and Francesco Di Filippo \at Center for Gravitational Physics, Yukawa Institute for Theoretical Physics, Kyoto University, Kyoto 606-8502, Japan, \email{francesco.difilippo@yukawa.kyoto-u.ac.jp}
\and
Stefano Liberati \at SISSA - International School for Advanced Studies, Via Bonomea 265, 34136 Trieste, Italy, \\IFPU - Institute for Fundamental Physics of the Universe, Via Beirut 2, 34014 Trieste, Italy and \\INFN - Sez. Trieste, Italy. \email{liberati@sissa.it}
\and
Matt Visser \at School of Mathematics and Statistics, Victoria University of Wellington, PO Box 600, Wellington 6140, New Zealand, \email{matt.visser@sms.vuw.ac.nz}
}
\maketitle
%====================================
\abstract{{Penrose's singularity theorem implies that if a  trapped region forms in a gravitational collapse, then a singularity must form as well within such region. 
However, it is widely expected that singularities should be generically avoided by quantum gravitational effects. 
Here we shall explore both the minimum requirements to avoid singularities in a gravitational collapse as well as discuss, without relying on a specific quantum gravity model, the possible regular spacetimes associated to such regularization of the spacetime fabric. 
In particular, we shall expose the intimate and quite subtle relationship between regular black holes, black bounces and their corresponding horizonless object limits. In doing so, we shall devote specific attention to those critical (extremal) black hole configurations lying at the boundary between horizonful and horizonless geometries.  
While these studies are carried out in stationary configurations, the presence of generic instabilities strongly suggest the need for considering more realistic time-dependent dynamical spacetimes. Missing specific dynamical models, much less rigorous statements can be made for evolving geometries. We shall nonetheless summarize here their present understanding and discuss their implications for future phenomenological studies.
}}

%\clearpage
%%%%%%%%%%%%%%%%%%%%%%%%%%%%%%%%%%%%%%%%%%%%%%%%%%%%%%%%%
%\tableofcontents %%% DRAFT ONLY
%%%%%%%%%%%%%%%%%%%%%%%%%%%%%%%%%%%%%%%%%%%%%%%%%%%%%%%%%

 %\clearpage
%\titlerunning{Singularity-free gravitational collapse}
%=============================================================
\section{Singularity regularization in effective geometries}
\label{sec:1}
%=============================================================

In this chapter we shall primarily investigate singularity regularization~\cite{Dymnikova:2003vt,Dymnikova:2004zc,Ansoldi:2008jw, Bardeen, Barcelo:2009tpa, Crowther:2021qij, Hayward:2005gi,Frolov:2014jva,Frolov:2016gwl,Frolov:2017rjz,Mazur:2001fv, Mazur:2004fk, Roman:1983zza} at a purely kinematical level; 
eschewing for now explicit use of the Einstein equations. The reasons for this are two-fold: First, even in standard general relativity, the Einstein equations~\cite{Einstein:1916vd} only have predictive power once you make (rather strong) assumptions on the nature of the stress-energy --- be it vacuum, or some nontrivial stress-energy satisfying some form of (semi-classical) energy condition~\cite{Barcelo:2002bv,  Borde:1987qr, Curiel:2014zba,Fewster:2010gm, Fewster:2012yh, Ford:1994bj,Ford:2003qt,Hochberg:1998ha, Kontou:2020bta,Mars:1996khm,Martin-Moruno:2017exc, Parker:1973qd,Yurtsever:1990gx,Zaslavskii:2010qz}. 
Second, if for some reason one wishes to step beyond standard general relativity, the status of the (modified) Einstein equations and (modified) energy conditions is even more fraught.
In view of these observations we shall see just how much we can do using purely kinematical observations.
Such questions are of considerable importance in view of recent dramatic advances in both observational techniques~\cite{LIGOScientific:2016aoc, LIGOScientific:2017vwq, LIGOScientific:2016sjg, LIGOScientific:2017ync, GRAVITY:2018ofz,EventHorizonTelescope:2019dse, EventHorizonTelescope:2019ths, EventHorizonTelescope:2019ggy} and phenomenological understanding~\cite{Bambi:2015kza,Berry:2020ntz, Broderick:2013rlq,Carballo-Rubio:2018jzw,Carballo-Rubio:2022imz,Cardoso:2016ryw, Gralla:2020pra, Held:2019xde,Johannsen:2015hib, Johannsen:2015mdd, Psaltis:2018xkc, Simpson:2021dyo, Simpson:2021zfl, Visser:2008rtf}. 

%=============================================================
\subsection{General relativity: singularity theorems and geodesic incompleteness}
\label{sec:1a}
%=============================================================

The existence of singularities is one of the most intriguing aspects of the theory of general relativity. From the conceptual subtleties in their definition~\cite{Geroch:1968ut} to the implicit suggestion of new physics that would avoid their formation~\cite{Garay:1994en}, singularities have been at the core of numerous developments in classical and quantum gravity~\cite{Crowther:2021qij}.

From a mathematical perspective, the formation of singularities in general relativity is unavoidable once certain conditions are met. These conditions are captured by the so-called singularity theorems~\cite{Penrose:1964wq,Hawking:1970zqf} (see also~\cite{Borde:1987qr,Fewster:2010gm,Ford:2003qt,Penrose:1969pc},  and see~\cite{Senovilla:2014gza} for a recent review).

From a physical perspective, these conditions are expected to be realized in two different kinds of astrophysical situations: the early universe (which results in the Big Bang singularity), and gravitational collapse (which results in black holes, or closely related objects). Here, we will be mostly concerned with the latter situation, adequately encapsulated in Penrose's singularity theorem~\cite{Penrose:1964wq}.

In a nutshell, Penrose's theorem demonstrates that,
once a closed trapped surface $\mathscr{S}^2$ is formed in a spacetime $\mathscr{M}$, there exists an incomplete geodesic in the causal future $J(\mathscr{S}^2)$ of $\mathscr{S}^2$. This result relies on a number of technical assumptions that are spelled out below. However, before diving into these assumptions, we think it can be useful to discuss in some detail the notions introduced above:

\begin{itemize}
\item 
A closed trapped surface $\mathscr{S}^2$ is a closed and spacelike 2-dimensional surface such that the area of all light fronts propagating through any of the points on the surface is decreasing toward the future. This is clearly associated with strong gravitational fields, as the standard behavior in weak gravitational fields (or in the absence of gravitational field) is that the area can either decrease or increase, depending on the initial conditions considered for light rays (the simplest example is that of exploding/imploding spheres of light in flat spacetime).
\item 
The causal future $J(A)$ of a set $A\subset\mathscr{M}$ is the set of points that are connected by past-directed causal (null and timelike) curves to the points in $A$. In other words, $J(A)$ is the set of points that can be reached following causal trajectories going through points in $A$.
\item 
An incomplete geodesic is a geodesic that cannot be followed indefinitely. For instance, an observer following a timelike incomplete geodesic will only be able to record its experience for a finite amount of proper time. A simple example of a geodesically incomplete spacetime, which illustrates how this notion captures the existence of ``holes'' in spacetime, is obtained by artificially removing a point from flat spacetime. (An incomplete null geodesic is one which terminates in finite affine parameter ``time''.)
\end{itemize}

Hence, the condition that must be met, according to Penrose's theorem, is the formation of closed trapped surfaces. The formation of closed trapped surfaces is reasonable from a conceptual perspective~\cite{Hawking:1973uf,Dafermos:2004wr,Andersson:2005gq}, and has also been reproduced numerically~\cite{Schnetter:2005ea}, so there is no plausible physical reason to doubt that this condition can be satisfied, at least within the framework of classical general relativity. 

Moreover, Penrose's theorem relies on the following assumptions (ordered in terms of increasing strength):
\begin{enumerate}
\item{The weak energy condition is satisfied.\label{cond1}}
\item{The Einstein field equations hold.\label{cond2}}
\item{Global hyperbolicity holds.\label{cond3}}
\item{Pseudo-Riemanniann geometry provides an adequate description of spacetime.\label{cond4}}
\end{enumerate}

Violating any of these assumptions would open up the possibility of getting rid of the singular behavior in Penrose's theorem. Hence, we can use this list of input assumptions to classify different approaches to this problem.

For instance, it is argued by many authors that spacetime must be fundamentally discrete --- a discretium rather than a continuum~\cite{Ambjorn:2000dv,Ambjorn:2001cv, Rovelli:1994ge, Sorkin:2003bx, Sotiriou:2011mu, Visser:2012best}. This would violate all of the above assumptions: being formulated in the mathematical framework of pseudo-Riemanniann geometry, they lose their meaning if the latter ceases to be applicable. It is therefore clear that fundamental discreteness can be one guiding principle leading to singularity-free theories. Nonetheless, we still lack a definitive theory of quantum gravity (albeit we do have some tentative calculations indicating that the resolution of singularities is indeed achieved~\cite{Ashtekar:2005qt}), so we shall take here a more humble approach. 

More specifically, our working framework will consist of hypothesising that both assumptions \{\ref{cond3}-\ref{cond4}\} hold, but we shall relax assumptions \{\ref{cond1}-\ref{cond2}\}. This is equivalent to assuming that pseudo-Riemanniann geometry provides a good description of the kinematics, while the only condition on the dynamics is that it leads to a well-posed initial value problem (hence the global hyperbolic condition).

Using the tools provided by pseudo-Riemanniann geometries has several practical advantages due to our familiarity with them~\cite{Penrose:techniques}, but we also believe this procedure has heuristic value from the perspective of understanding the main features to be expected in a framework that supersedes general relativity. 

Basically, we are entailing scenarios in which, after the formation of a trapped region, a quantum gravitational description is circumscribed to some finite region of spacetime (possibly associated with Planckian densities). The outcome of this evolution is then a globally hyperbolic geometry, regular and classical everywhere.\footnote{This can be relaxed so to admit Planck scale regions still requiring a quantum gravitational description, as long as these are not considered as missing points from the manifold, as indeed in this case one can still consider to cover them by analytically extending the regular geometry describing the rest of the spacetime.}

%======================================
\subsection{Beyond general relativity: deforming black holes into geodesically complete spacetimes}
\label{sec:1b}
%=====================================

In this section, we discuss the different kinds of geometries describing singularity-free black holes. Our precise definition of these geometries is based on the following features
\begin{itemize}
\item 
Global hyperbolicity.
\item 
Geodesic completeness.
\item 
Asymptotic flatness.
\item 
Existence of a closed trapped surface $\mathscr{S}^2$ (the boundary of the black hole).
\item 
Finiteness of the curvature invariants.
\end{itemize}
As explained in the previous sections, these geometries must violate at least one of the assumptions \{\ref{cond1}-\ref{cond2}\}. In this sense, Penrose's theorem is a convenient starting point. 

We can begin by noticing that in the latter the existence of incomplete geodesics is intimately linked to the formation of a focusing point (defined as a point where a congruence of geodesics is characterized by a vanishing cross-sectional area).

Let us make some of these notions more mathematically precise. We are assuming the existence of a spacelike trapped surface $\mathscr{S}^2$, which can be defined using the two null vector fields that are normal to it. The (3+1) dimensionality of $\mathscr{M}$ and 2 dimensionality of $\mathscr{S}^2$, together with the spacelike character of the latter, imply that there are two linearly independent (future-directed) normal null vectors at each point of $\mathscr{S}^2$. We will call these two independent normal null vectors $\bm{l}$ (outgoing null normal) and $\bm{k}$ (ingoing null normal). If we define $h_{ab}$ as the 2-metric induced  on $\mathscr{S}^2$, the expansions along these vector fields are given by:
\begin{equation}
\theta^{(\bm{X})}={1\over\sqrt{h}} \; \mathcal{L}_{\bm{X}}\sqrt{h}=h^{ab}\nabla_a X_b,\qquad \bm{X}\in\{\bm{l},\bm{k}\},
\end{equation}
where $\mathcal{L}_{\bm{X}}$ is the Lie derivative along $\bm{X}$, $h=\mbox{det}(h_{ab}$) and $\bm{\nabla}$ is the 4-dimensional covariant derivative. 

According to the definition above, the expansion $\theta^{(\bm{X})}$ measures the local change in the area of $\mathscr{S}^2$ when the latter undergoes a local deformation along the vector field $\bm{X}$. 
The expansion $\theta^{(\bm{k})}$ being negative is the standard behavior expected for ingoing geodesics, while $\theta^{(\bm{l})}$ is in a flat spacetime  (or a spacetime describing weak gravitational fields) always positive.  
A trapped surface is defined in terms of these expansions as:
\begin{equation}
\theta^{(\bm{k})}<0,\qquad\qquad \theta^{(\bm{l})}<0.
\end{equation}
and a focusing point is characterized by the divergence of the outgoing expansion: $\theta^{(\bm{l})}\to-\infty$

The proof of Penrose's theorem goes schematically as follows: First the Raychaudhuri equation is used to prove that the Einstein field equations and the null energy conditions imply that $\theta^{(\bm{l})}$ keeps becoming more negative towards the future, until reaching a focusing point at finite affine parameter $\lambda=\lambda_0$ at which $\left.\theta^{(\bm{l})}\right|_{\lambda=\lambda_0}=-\infty$ for all null geodesics. On the other hand, purely geometrical arguments can be used to show that the existence of such focusing point is incompatible with the existence of a non-compact Cauchy hypersurface\footnote{No assumptions on the topology of the Cauchy hypersurface are actually required if we instead assume the existence of at least one geodesic that does not fall into the black hole \cite{Hawking:1973uf}.}.

\begin{figure}[!htbp]%
\begin{center}
\vbox{\includegraphics[width=0.9\textwidth]{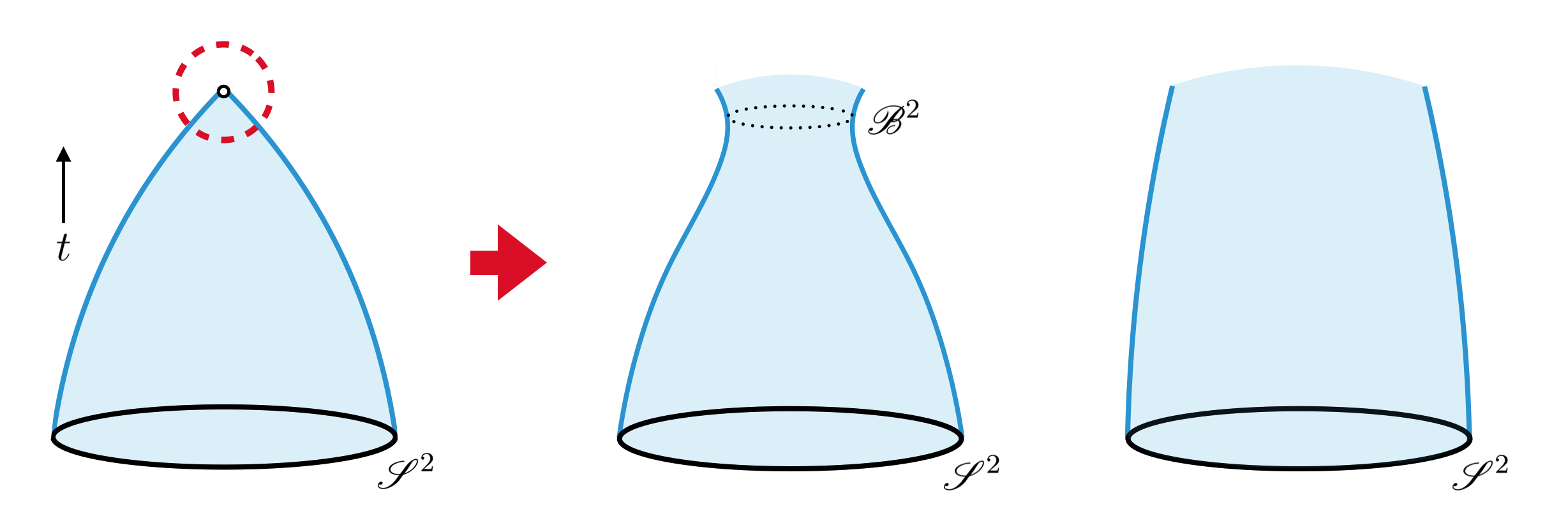}}
\bigskip%
\caption{Avoiding that the spacetime be geodesically incomplete entails modifying the geometry in the surroundings of the focusing point in Penrose's theorem, in such a way that either a defocusing point is created --- either at a finite affine distance (thus also creating the 2-surface $\mathscr{B}^2$ displayed above) or at infinite affine distance --- or the focusing point is displaced to infinite affine distance. The figure on the right is compatible with the two last cases. Ingoing radial null geodesics are not included in this picture as, for each of these cases, these geodesics can display different behaviors that are analyzed in detail in the text.}
\label{fig:possibilities}%
\end{center}
\end{figure}

Thus, following Penrose's theorem, in order to make the spacetime geodesically complete we need to modify its geometry in the vicinity of the focusing point, either by creating a defocusing point or by displacing the focusing point to infinite affine distance (see Fig. \ref{fig:possibilities}). Equivalently, the expansion $\theta^{(\bm{l)}}(\lambda)$ must remain finite for all the possible values of $\lambda\in[0,\infty)$ (where we are identifying $\lambda=0$ with $\mathscr{S}^2$ without loss of generality).
%------------------------------
%------------------------------

These considerations constrain the behavior of outgoing null geodesics: the outgoing expansion can either remain negative but finite (thus having no defocusing points), vanish asymptotically for infinite affine distance without a defocusing point, vanish asymptotically for infinite affine distance with a defocusing point, or vanish at a finite affine distance (thus having a defocusing point at a finite affine distance). Simultaneously, ingoing null geodesics can generally display two qualitatively different behaviors, namely either being negative or being non-negative at the defocusing point. Combining these possibilities shows that the number of possible qualitatively different behaviors around the defocusing point is eight.

However, not all these geometries are regular. If we restrict for simplicity to the spherically symmetric case (no symmetry was required in the discussion up to this point), we can show that there are only four regular classes. The technical details are discussed in Ref.~\cite{Carballo-Rubio:2019fnb}, in which the four regular classes of geometries have been identified to be:
\begin{enumerate}
\item 
\textit{Evanescent horizon. The singularity is replaced by an inner horizon at which the $\theta^{(\bm{l)}}$ changes sign while $\theta^{(\bm{k)}}$ stays negative. The inner and the outer horizon merge in a finite time. This purely geometrical classification is blind to the type of the dynamical process or the timescale involved.} \label{class1}
\item 
\textit{Hidden wormhole. The singularity is replaced by a (global or local) minimum radius hypersurface. Both expansions change signs. Such a structure is reminiscent of a wormhole throat hidden inside a trapping horizon.} \label{class2}
\item 
\textit{Everlasting horizon. As in the evanescent horizon case, the geometries in this class possess an inner and an outer horizon. In this case, the two horizons never merge. This class can be seen as the limit of the evanescent horizon class when the timescale of merging of the two horizons is pushed out to infinity.} \label{class3}
\item 
\textit{Asymptotic hidden wormhole. The singularity is replaced by a global minimum radius hypersurface that is reached in an infinite affine time. This class can be obtained from the hidden wormhole class by pushing the wormhole throat out to an infinite affine distance.} \label{class4}
\end{enumerate}
There are two main aspects about these geometries that are essential for our discussion  below:
\begin{itemize}
    \item The spacetimes of interest for us describe the collapse of a regular distribution of matter from a given initial Cauchy surface with topology $\mathbb{R}^3$. Any geometry that satisfies this condition and belong to one of the classes above must be dynamical.    
    \item Classes \{\ref{class1}, \ref{class3}\} above are simply connected, and differ in their dynamical behavior only. On the other hand, classes \{\ref{class2}, \ref{class4}\} above are non-simply connected. 
\end{itemize}
Instead of working directly with time-dependent situations, we will start considering static situations and focus on the second property above, namely whether the regularization mechanism results in either simply or non-simply connected spacetimes. 

After discussing what static spacetimes with these different topologies look like, we will discuss time-dependent situations. This provides an intuitive and gradual way of understanding the main differences between these classes.

%==========================================
\section{Geodesically complete alternatives to static black holes}
\label{sec:2}
%==========================================

As mentioned above, we will start our discussion with static situations, namely with no explicit time dependence in the metric. The reader must keep in mind that the geometries discussed in this section have therefore no relevance for the discussion of gravitational collapse, though they provide a stepping stone towards constructing these geometries, as we will discuss in more detail below.

The most generic static spherically symmetric line element is given in this case by
\begin{equation}
    ds^2=-F(r)\,dt^2+F^{-1}(r)\,dr^2+\rho^2(r)\,d\Omega^2\,,
\end{equation}
where $d\Omega^2$ is the usual line element on the 2-sphere. (These are sometimes called Buchdahl coordinates~\cite{Boonserm:2007zm,Lobo:2020ffi}.) We can also use the Eddington-Finkelstein form
\begin{equation}
    ds^2=-F(r)\,dv^2+2\,dv\,dr+\rho^2(r)\,d\Omega^2\,.
\end{equation}
The line element has a trapping horizon at $r=r_{\rm H}$ whenever $F(r_{\rm H})=0$ vanishes. In fact, one can show explicitly that:
\begin{equation}
    \theta^{(\bm{l})}= 2F(r)\;{\partial_r\rho(r)\over {\rho(r)}}; 
    \qquad\hbox{and}\qquad
    \theta^{(\bm{k})}= -2\;{\partial_r\rho(r)\over \rho(r)},
\end{equation}
where we have normalized $g(\bm{l},\bm{k})=-2$.

This trapping horizon can be~\cite{Hayward:1993wb}:
\begin{itemize}
\item \textit{inner} if $F'(r_{\rm H})<0$,
\item \textit{outer} if $F'(r_{\rm H})>0$.
\end{itemize}
We know that geometries that have an outer trapping horizon and satisfy the assumptions in Penrose's theorem are singular. An example is the Schwarzschild metric
\begin{equation}
    F(r)=1-\frac{2M}{r}\,,\qquad\qquad \rho^2(r)=r^2\,,
\end{equation}
that becomes singular in the limit $r\rightarrow0$.

In order to avoid scalar curvature singularities, either $F(r)$ has an even number of zeros, or $\rho(r)$ has a minimum~\cite{Carballo-Rubio:2019fnb}. It is clear that none of these conditions is satisfied by the Schwarzschild metric above. However, we can consider a specific deformation of the Schwarzschild geometry satisfying both criteria for singularity regularization:
\begin{equation}\label{eq:example}
    F(r)=1-\frac{2M\rho^2(r)}{\rho^3(r)+2M\ell_1^2}\,,\qquad\qquad \rho^2(r)=r^2+\ell_2^2\,.
\end{equation}
This example has the following interesting features:
\begin{itemize}
\item 
It reduces to the so called Simpson--Visser metric for $\ell_1=0$ (see~\cite{Simpson:2018tsi, Simpson:2019mud, Lobo:2020ffi}, see also  \cite{Franzin:2021vnj} and \cite{Mazza:2021rgq} for its extension to rotating configurations). There is then no singularity due to the existence of a wormhole throat, with a radius proportional to $\ell_2$. Whether or not there is an outer horizon depends on the value of $\ell_2$ with respect to $M$.
\item 
It reduces to the Hayward metric for $\ell_2=0$ \cite{Hayward:2005gi}. There is then no singularity due to the existence of an even number of horizons, with an inner horizon radius proportional to $\ell_1.$ Whether or not there is an outer horizon (and thus an accompanying inner horizon) depends on the value of $\ell_1$ with respect to $M$.
\item 
It reduces to the (singular) Schwarzschild metric for $\ell_1=\ell_2=0$.
\end{itemize}
In this simple example, we see something that is generic: as anticipated by our local analysis of the defocusing point, there are two qualitatively distinct regularization mechanisms,  characterized by the topology of the resulting spacetimes:
\begin{itemize}
\item A regularization mechanism that is based in the introduction of at least one inner horizon ($\ell_1\neq0$), resulting in simply connected spacetimes. 
\item A regularization mechanisms based on the introduction of a wormhole throat ($\ell_2\neq0$), which produces non-simply connected spacetimes. 
\end{itemize}
This observation continues to hold true for more general static geometries, as well as for time-dependent situations.

For the sake of simplicity, in the following we will continue working with time-independent situations, and we will moreover focus on the example introduced in Eq.~(\ref{eq:example}). This will keep the geometries simple enough to be tractable without losing sight of any of the interesting physics. The reader should keep in mind that any of the 4 generic classes above can be reconstructed by considering a sequence of the model spacetimes of Eq.~(\ref{eq:example}) in which $M$, $\ell_1$ and $\ell_2$ become functions of time. We will discuss some features of such dynamical spacetimes below. 

%==========================================================
\subsection{Regularization in simply connected topologies}

%==========================================================

Within the family of geometries we are considering as an example, see Eq.~(\ref{eq:example}), the simply connected topologies are given by $\ell_2=0$, so that we have the metric
\begin{equation}
\label{E:simple}
    F(r)=1-\frac{2M\rho^2(r)}{\rho^3(r)+2M\ell_1^2}\,,\qquad\qquad \rho^2(r)=r^2.
\end{equation}
This metric can describe three distinct kinds of objects, depending on the relative values of the parameters $M$ and $\ell_1$ controlling the roots of the polynomial in the numerator of $F(r)$:
\begin{equation}
    F(r)=\frac{r^3-2Mr^2+2M\ell_1^2}{r^3+2M\ell_1^2}\,.
\end{equation}
The properties of these different objects are described in the sections below. We will keep $M$ fixed and explore this family of geometries as $\ell_1$ takes values in the interval $(0,+\infty)$.

%===========================================
\subsubsection{Regular black holes}
%===========================================
\label{sec:Reg-simply1}

\begin{figure}[htbp!]
    \centering
    \includegraphics[width=8cm]{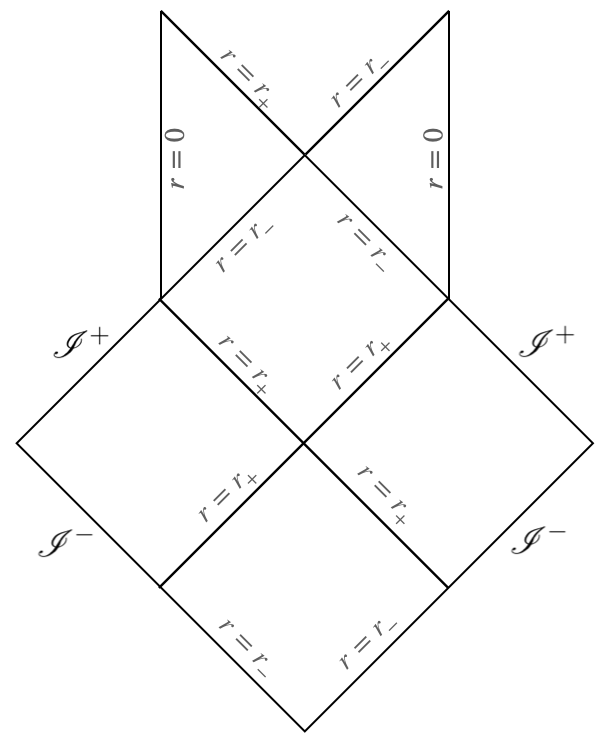}
    \caption{Two-horizon RBH, of the type presented in Eq.~(\ref{E:simple}), corresponding to $0<\ell_1<\ell_1^\star\equiv 4M/(3\sqrt{3})$. For the maximally extended Penrose diagram one should repeat the construction infinitely many times in the vertical direction. Note that the Penrose diagram is qualitatively similar to that for Reissner--Nordstr\"om --- except that the timelike curve $r=0$ is now carefully arranged to be regular, not singular. }
    \label{F:rbh2}
\end{figure}

For $\ell_1$ sufficiently small, 
more precisely $\ell_1\in(0,\ell_1^\star\equiv 4M/[3\sqrt{3}])$, the function $F(r)$ has two distinct positive roots $ r_\pm $ which correspond to outer and inner horizons as defined above. (The third negative root is in this situation unphysical.)
See Fig.~\ref{F:rbh2} for the relevant Penrose diagram. The derivative of $F(r)$ at each of the horizons provides their surface gravities $\kappa_\pm$,
\begin{equation}
    \kappa_\pm=\left.\frac{1}{2}\frac{dF}{dr}\right|_{r=r_\pm}\,,
\end{equation}
and, according to the definition of outer and inner horizons, we have 
\begin{equation}
    \kappa_-<0\,,\qquad\qquad \kappa_+>0\,.
\end{equation}
This implies that at the inner horizon there is a exponential focusing of null rays~\cite{Carballo-Rubio:2018pmi, Carballo-Rubio:2021bpr, Carballo-Rubio:2022twq, DiFilippo:2022qkl}. Within general relativity, it is well known that such focusing leads to an exponential instability of the inner horizon \cite{Poisson:1990eh} with a characteristic timescale fixed by the inner horizon surface gravity $\kappa_-$. 

In the context of regular black hole geometries, we do not know the dynamics of the theory. However, it is still possible to prove a linear instability on purely geometrical grounds, once a small perturbation is added to the background~\cite{Carballo-Rubio:2018pmi,Carballo-Rubio:2021bpr,DiFilippo:2022qkl, Carballo-Rubio:2022twq}. While it is impossible determine the endpoint of such instability without knowing the field equations of the theory, the presence of a linear instability constitutes a very general result --- one that should be taken as a strong cautionary note regarding the viability, as a stable final state, of any geometry with a generic inner horizon. 

The geometry around the inner horizon can be deformed so that $\kappa_-=0$, which removes the unstable behavior~\cite{Carballo-Rubio:2022kad,Franzin:2022wai}. These \textit{inner-extremal} regular black holes may therefore represent a suitable end state towards which the dynamical evolution triggered by the unstable nature of generic inner horizons could tend to. However, this problem has not been analyzed in detail yet, and other possibilities remain open.
%%%%%%%%%%%%%%%%%%%%%%%%%%%%%%%%%%%%%%%%%%%%%%%%%%
%%%%%%%%%%%%%%%%%%%%%%%%%%%%%%%%%%%%%%%%%%%%%%%%%%%
%===========================================
\subsubsection{Extremal regular black holes}
%===========================================
\label{sec:Reg-simply2}

\begin{figure}
    \centering
    \includegraphics[width=6cm]{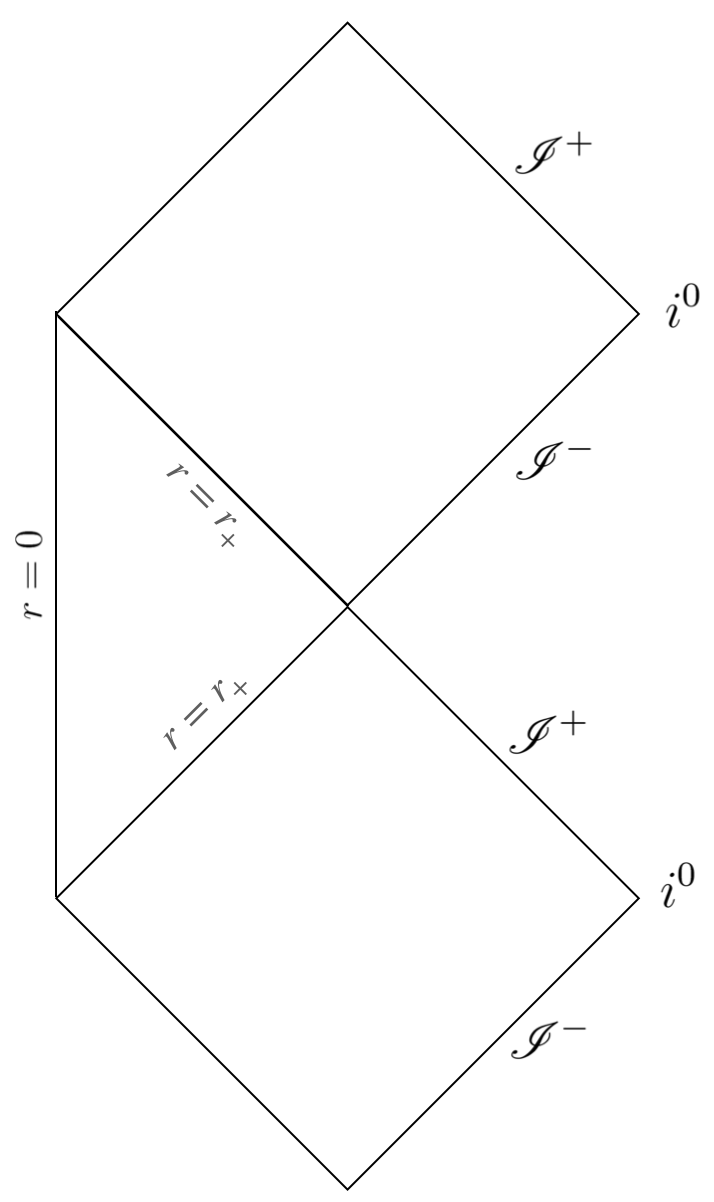}
    \caption{Extremal regular black hole, of type presented in Eq.~(\ref{E:simple}), corresponding to $\ell_1=\ell_1^\star$. As for the non-extremal case, the maximally extended Penrose diagram is obtained repeating the construction infinitely many times in the vertical direction, and it is qualitatively similar to that for an extremal Reissner--Nordstr\"om black hole --- except that the timelike curve $r=0$ is now carefully arranged to be regular, not singular. }
    \label{fig:extr_rbh}
\end{figure}

\begin{figure}
    \centering
    \includegraphics[width=9cm]{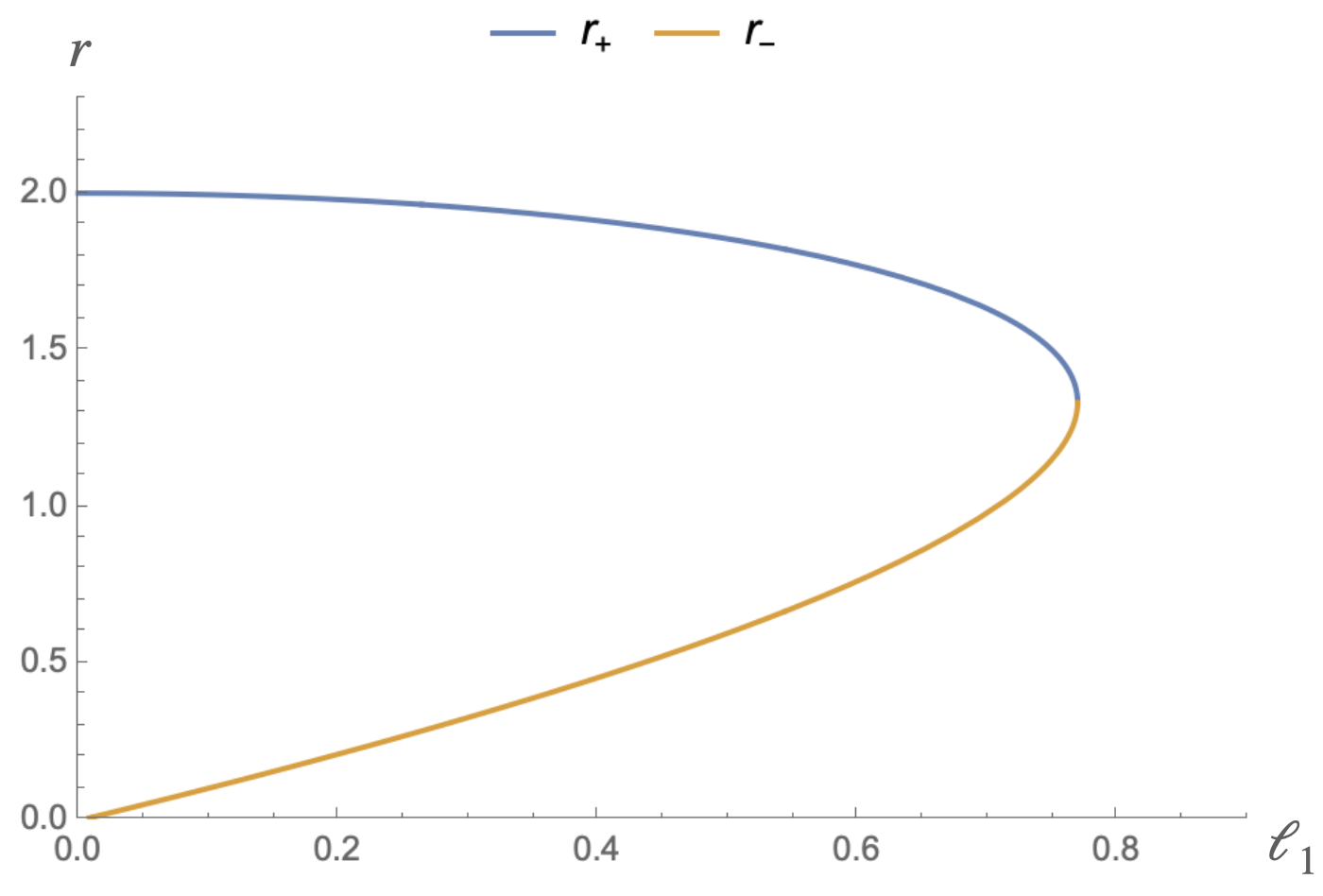}
    \caption{{Location of outer and inner horizons of the geometry specified in (\ref{E:simple}), as determined by the roots of $F(r)=0$. This is equivalent to determining the positive roots of $r^3-2Mr^2+2M\ell_1^2$. Note that the two horizons merge, to form a single extremal horizon, when $\ell_1\to \ell_1^\star\equiv 4M/[3\sqrt{3}] \approx 0.77 M $. When this happens one has $r_E=4M/3 \approx 1.333 M.$}}
    \label{fig:RBH}
\end{figure}

As seen in the Penrose diagram presented in Fig.~\ref{fig:RBH}, if we increase the value of the parameter $\ell_1$, the outer and inner horizon move towards each other. In particular, for $\ell_1\to\ell_1^\star\equiv 4M/(3\sqrt{3})$, the  two horizons merge into a single extremal horizon, located at $r_E=4M/3$, and the corresponding spacetime describes an extremal regular black hole.
(There is also a third unphysical root at $r_{\rm unphysical}=-2M/3$.)
Fig.~\ref{fig:extr_rbh} shows the Penrose diagram corresponding to this configuration

Specifically in this extremal limit
\begin{equation}
    r^3 - 2M(r^2-\ell_1^2) \to r^3 - 2M(r^2-[\ell_1^*]^2) 
    = {(3r+2M)(3r-4M)^2\over 27}.
\end{equation}

\noindent
Geometrically it is guaranteed that very special  things happen at all extremal horizons; a fully general analysis is presented in Appendix 1. For now let us just observe that in the current context outside the outer horizon, and inside the inner horizon, the Misner--Sharp quasilocal mass must satisfy $2m(r) < r$. Thence as inner and outer horizons merge at an extremal $r_E$ we must have both $2m'(r_E)=1$ and $m''(r_E) < 0$. Furthermore on the extremal-horizon curvature invariants are functions only of $r_E$ and $m''(r_e)$. These observations survive even in situations much more general than the current 1-parameter modification of Schwarzschild ($\ell_1\neq0$, $\ell_2=0$). See Appendix 1 for details.

%%%%%%%%%%%%%%%%%%%%%%
%%%%%%%%%%%%%%%%%%%%%%%

%===========================================
\subsubsection{Horizonless objects}
%===========================================
\label{sec:Reg-simply3}

Geometries for which $\ell_1\in(\ell_1^\star= 4M/[3\sqrt{3}],+\infty)$ do not contain any type of horizon. We will call these \textit{horizonless stars}, as these geometries describe a spherical distribution of matter surrounded by vacuum. These represent another possible end state for the dynamical evolution of regular black holes.

Depending on the value of $\ell_1$, these stars are more or less compact. Indeed for $\ell_1-\ell_1^\star \ll \ell_1^\star$, the corresponding stars are ultracompact, while in the opposite limit $\ell_1\rightarrow+\infty$ the stars become more and more dilute until eventually becoming indistinguishable from flat spacetime. Regarding the causal structure of these spacetimes, the Penrose diagram for any of these geometries is equivalent to that of Minkowski spacetime.

Concerning the formation of such horizonless stars, it appears far from obvious that they can be realized without an intermediates state involving a trapped region. Indeed, without exotic physics entering at energy densities beyond nuclear density, it is difficult to avoid the conclusion that the gravitational collapse will occur almost in free fall~\cite{Barcelo:2007yk,Barcelo:2009zz,Barcelo:2009tpa}. In this case, it was shown that even quantum effects cannot change the expected behaviour and prevent the formation of a trapping horizon~\cite{Barcelo:2007yk,Barcelo:2009zz,Barcelo:2009tpa}.\footnote{This can be somewhat expected based on the heuristic argument that, if ones starts at early times with a dilute star in quasi-Minkoswki vacuum then, in the absence of a Cauchy horizon~\cite{Fulling:1978ht}, a free-fall collapse would allow one to keep renormalizing the stress energy tensor at different radii, obtaining small deviations from the initial vacuum in the local inertial frame. Hence, this would prevent the build up of large quantum effects able to slow down the collapse before the formation of a trapping horizon. Of course, one might consider the possibility of different initial conditions, e.g.~concerning the vacuum state at past null infinity, something that has so far received quite limited attention (see e.g.~\cite{Fredenhagen:1989kr}).} However, the above mentioned, generic inner horizon instability of regular black holes seems to entail a possible evolution of regular black holes towards some sort of stable configuration. This could be for example an inner extremal regular black hole with $\kappa_-=0$ at the inner horizon \cite{Carballo-Rubio:2022kad,Franzin:2022wai}, or possibly the ultracompact but horizonless limit of the very same geometric family~\cite{Carballo-Rubio:2022nuj}.

Concerning the stability, not very much can be safely said due to the lack of a proper understanding of the dynamics associated to these objects. Nonetheless it is interesting to note that horizonless stars are generically endowed with pairs of light rings (closed photon orbits): an outer, unstable, one corresponding to the usual structure present e.g.~in Schwarzschild spacetime, and an inner, stable one (at least for static configurations)~\cite{Cunha:2022gde,Hod:2022mys, Bargueno:2022vkf}. This novel feature is potentially dangerous as it might lead to a nonlinear instability due to the accumulation of energy (e.g.~photons and/or gravitons) which might bring the horizonless star back within its gravitational radius and hence lead to the formation of a trapping horizon (see e.g.~ the discussion of this instability for the case of boson stars~\cite{Cunha:2022gde}). While this is surely a feature worth investigating in greater detail (see e.g. the caveats raised in~\cite{Zhong:2022jke}), for now  we just want to point out that such inner light ring necessarily has to appear in a region where the metric has order one deviations from the Schwarzschild one, which is tantamount to saying that they will lie in non vacuum regions. (Alternatively one might say that inner light rings will always lie below the stellar surface if this is defined as containing almost all of the ADM mass.) In this sense any study concerning the potential instability at these inner light rings cannot avoid the need to  model the matter interaction with the massless field --- such as for example its reflectivity or absorption properties --- which would generally dampen this possible unstable behavior.

%================================================================
\subsection{{Regularization in non-simply connected topologies}}
%================================================================

Within the family of geometries we are considering as an example, the non-simply connected topologies are given by $\ell_1=0$, so that we have the Simpson--Visser metric
\begin{equation}
    F(r)=1-\frac{2M}{\rho(r)}\,,\qquad\qquad \rho^2(r)=r^2+\ell_2^2.
\end{equation}
The geometries in this class have the characteristic of possessing a minimum areal radius $\rho_{\rm min}=\rho(0)=\ell_2$. This metric can describe three distinct kinds of objects, depending on the relative values of the parameters $M$ and $\ell_2$ controlling the roots of the polynomial in the numerator of $F(r)$:
\begin{equation}
    F(r)=\frac{\sqrt{r^2+\ell_2^2}-2M}{\rho(r)}\,.
\end{equation}
The properties of these different objects are described in the sections below. We will keep $M$ fixed and explore this family of geometries as $\ell_2$ takes values in the interval $(0,+\infty)$.

%\bigskip
%\noindent
%\red{\bf Note actual error in the previous version of equation above; \\
%this propagates below... \\
%Check by dimensional analysis... --- M}

%===========================================
\subsubsection{Hidden wormholes}
%===========================================
\label{SS:hidden}
%===========================================

For $\ell_2$ sufficiently small, more precisely $\ell_2\in(0,\ell_2^\star\equiv 2M)$, the function $F(r)$ has a single 
 (positive) root $ r_+$ which corresponds to an outer horizon, similarly to the cases discussed above.
The location of this root is given by the condition
\begin{equation}\label{eq:trap_reg}
r^2=4M^2-\ell^2_2\,.
\end{equation}
That is 
\begin{equation}\label{eq:trap_reg2}
r_\pm = \pm \sqrt{4M^2-\ell_2^2}.
\end{equation}

The regularization is achieved in this case, not by the introduction of an inner horizon, but by the introduction of a minimum length $\ell_2$ so that spheres in this spacetime cannot have an area below $4\pi(\ell_2)^2$. This implies that the topology of these spacetimes is $\mathbb{R}^2\times S^2$, which are therefore non-simply connected. The minimum radius hypersurface is within the trapped region and it is spacelike. Hence, it can only by traversed in one direction.

\begin{figure}[htbp!]
    \centering
    \includegraphics[width=8cm]{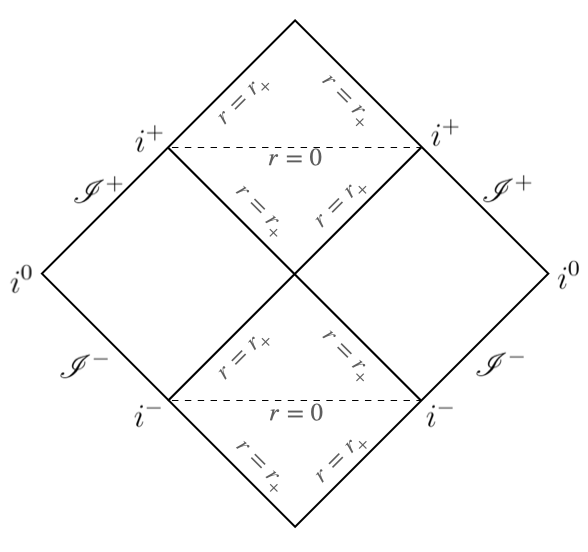}
    \caption{``Black bounce'': In this situation the (spacelike) wormhole throat at $r=0$ is a regular part of the spacetime and is hidden behind a horizon. For the maximally extended Penrose diagram one should repeat the construction infinitely many times in the vertical direction. (One could attempt to ``simplify'' the Penrose diagram by identifying past and future wormhole throats; this would indeed make the maximally extended Penrose diagram simpler, at the cost of introducing closed timelike curves (CTCs), which in particular would destroy global hyperbolicity.) }
    \label{fig:splk}
\end{figure}

From a physical perspective, these spacetimes describe black holes that contain a wormhole throat in their interior. 
The wormhole throat in this particular situation is a spacelike hypersurface which is ``hidden'' for observers that remain outside the black hole. 
This is typically referred to as a ``black bounce''~\cite{Simpson:2018tsi, Simpson:2019cer}. 
One horizon is a black hole horizon in ``our" universe, the other is a white hole horizon in the ``future'' universe. 
See Fig.~\ref{fig:splk}  for a suitable Penrose diagram. The maximally extended Penrose diagram for the ``black bounce'' spacetime is qualitatively similar to an infinite vertical stack of Schwarzschild spacetime Penrose diagrams --- except that what was the spacelike singularity at $r=0$ has been replaced by a regular spacelike bounce into the next part of the Penrose diagram. The main issue with these geometries is that, for these to arise in gravitational collapse with standard initial conditions, there must be a change of topology of Cauchy hypersurfaces from $\mathbb{R}^3$ to $\mathbb{R}\times S^2$. This is incompatible with global hyperbolicity~\cite{Bernal:2003jb}.

%%%%%%%%%%%%%%%%%%%
%%%%%%%%%%%%%%%%%%%%%%

%===========================================
\subsubsection{Null wormholes}
%===========================================

\begin{figure}[htbp!]
    \centering
    \includegraphics[width=7cm]{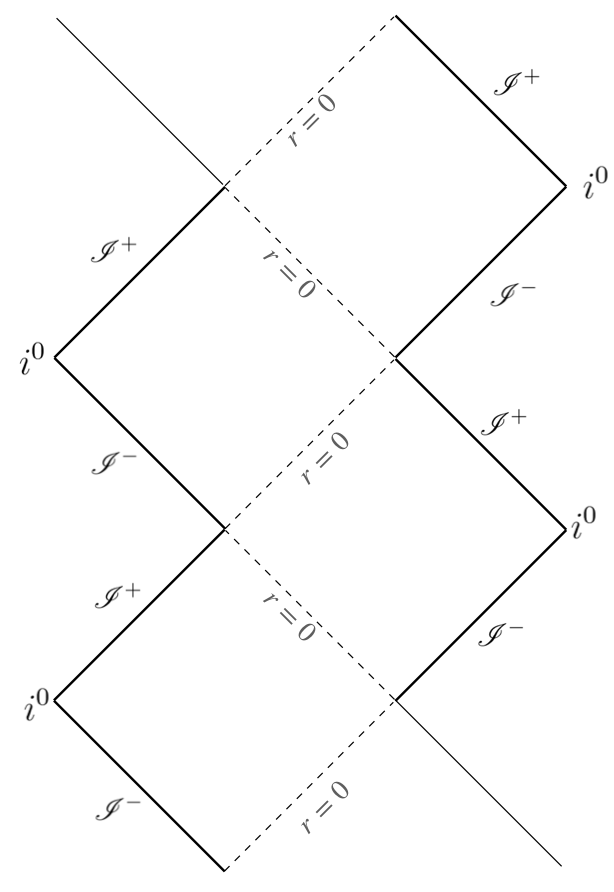}
    \caption{Null wormhole throat. For the maximally extended Penrose diagram one should again repeat the construction infinitely many times in the vertical direction. (One could again attempt to ``simplify'' the Penrose diagram by identifying past and future null wormhole throats; this would indeed make the maximally extended Penrose diagram simpler, at the cost of introducing closed timelike curves (CTCs), which in particular would destroy global hyperbolicity.)}
    \label{fig:null}
\end{figure}

As seen in the Penrose diagram of Fig.~\ref{fig:null}, if we now increase the value of the parameter $\ell_2$, the wormhole throat at $r=0$ and the horizons at $r= \pm ([2M]^2-\ell_2^2)^{1/2}$ move towards each other. In particular, for $\ell_2\to\ell_2^\star\equiv 2M$, both structures merge into a null wormhole throat. Because the wormhole throat is null, it is at best one-way traversable~\cite{Simpson:2018tsi, Simpson:2019cer}. These null throat wormholes represent a new paradigm that lies well outside the traditional realm of two-way traversable wormholes such as those discussed in~\cite{Morris:1988cz,Morris:1988tu, Visser:1989kh, Visser:1989kg, Visser:1995cc} and~\cite{Boonserm:2018orb, Hochberg:1997wp, Hochberg:1998ha, Lobo:2020ffi, Simpson:2018tsi, Simpson:2019cer}.

%%%%%%%%%%%%%%%%%%%%%%%%%%%%%%%%%
%%%%%%%%%%%%%%%%%%%%%%%%%%%%%%%%%

%===========================================
\subsubsection{Naked wormholes}
%===========================================

\begin{figure}[htbp!]
    \centering
    \includegraphics[width=6cm]{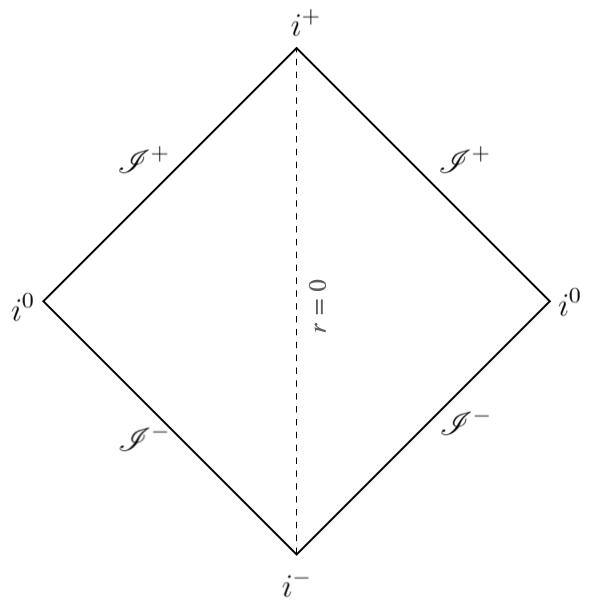}
    \caption{Timelike wormhole throat at $r=0$. This implies a ``naked'', in principle globally traversable,  wormhole. To determine practical traversability one would additionally need to ensure that tidal forces could be kept suitably small. }
    \label{F:timelike_throat}
\end{figure}

Geometries for which $\ell_2\in(\ell_2^\star\equiv 2M,+\infty)$ do not contain any type of horizon. 
See Fig.~\ref{F:timelike_throat} for a suitable Penrose digram. 
In these geometries, the wormhole throat at $r=0$ is now timelike, and is no longer ``hidden''. Such wormholes would be (at least in principle) globally traversable, and have been the subject of intensive investigation for quite different reasons. See for instance~\cite{Morris:1988cz,
Morris:1988tu, Visser:1989kh, Visser:1989kg, Visser:1995cc} and~\cite{Boonserm:2018orb, Hochberg:1997wp, Hochberg:1998ha, Lobo:2020ffi, Simpson:2018tsi, Simpson:2019cer}.
The qualitative analysis presented herein does not extend to quantitatively estimating the tidal forces, which would be necessary for verifying practical traversability.

%==========================================================
\subsection{Comment on ``mixed" geometries
}
%==========================================================
For completeness, let us consider the case in which both $\ell_1$ and $\ell_2$ are different from zero. Within the family of geometries we are considering as an example, we have the metric
\begin{equation}
   F(r) = 1 - {2M \rho^2(r)\over \rho^3(r) + 2 M \ell_1^2}\,,
   \qquad\qquad \rho^2(r)=r^2+\ell_2^2.
\end{equation}
This metric can describe three distinct kinds of objects, depending on the relative values of the parameters $M$ and $\ell_1$ controlling the roots of the polynomial in the numerator of $F(r)$:
\begin{equation}
    F(r)=\frac{\rho^3(r)-2M(\rho^2(r)-\ell_1^2)}{\rho^3(r)+2M\ell_1^2}\,.
\end{equation}
The properties of these different objects are sketched below.

Similarly to what we saw with previous examples, horizons are located at the roots of the polynomial
\begin{equation}
    P(\rho) \equiv \rho^3 - 2 M (\rho^2-\ell_1^2). 
\end{equation}
This is a cubic in $\rho$, so mathematically has three roots $\rho_i(M,\ell_1)$ with $i\in\{1,2,3\}$, not all of which need be physical. This then corresponds to six roots in terms of $r$, symmetrically placed around $r=0$, at locations:
\begin{equation}
    r_i(M, \ell_1,\ell_2) = \pm \sqrt{ \rho_i^2(M,\ell_1)-\ell_2^2} . 
\end{equation}
Not all of these roots need be real, which limits the number of actual horizons around the wormhole throat.

To find the three roots $\rho_i(M,\ell_1)$
the analysis is then very similar to that for the $\ell_2=0$ special case analyzed above.
We again define 
\begin{equation}
    \ell_1^\star = {4M\over3\sqrt 3}.
\end{equation}
For $\ell< \ell_1^\star$ there are three real roots $\rho_i(M,\ell_1)$, two of them positive and one negative. 
For $\ell= \ell_1^\star$ there are two real roots, a repeated positive root at $\rho=4M/3$, and a singleton negative root at $\rho=-2M/3$.
For $\ell > \ell_1^\star$ there is only one real root,  a singleton negative root. Note that, in order to keep the physical interpretation of $\rho$ as an areal radius, as well as for recovering the standard Hayward regular black hole solution in the limit $\ell_2\to 0$, one needs to take $\rho>0$ and hence keep only the positive roots in the above analysis.

\begin{figure}
    \centering
    \includegraphics[width=5cm]{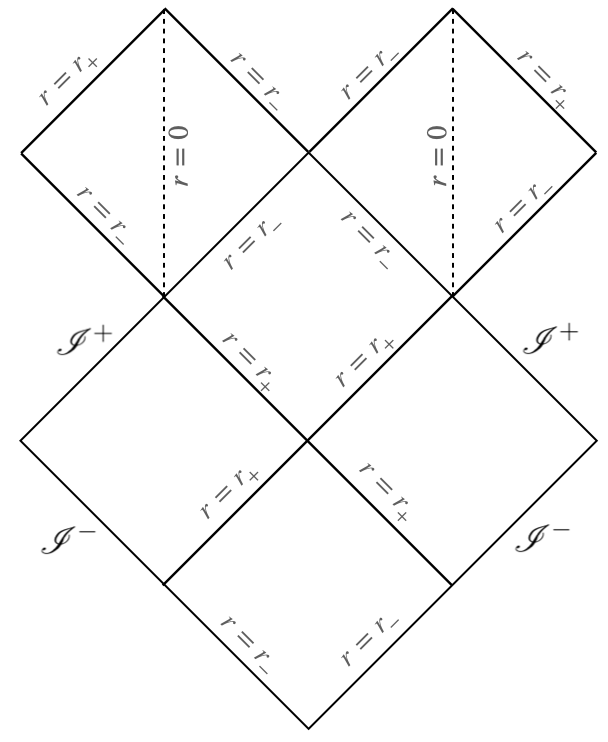}
    \caption{Timelike wormhole throat inside the inner horizon obtained for $\ell_1\neq 0$ and $\ell_2$ sufficiently small. The full Penrose diagram is obtained by repeating the construction both vertically and horizontally. }
    \label{fig:tmlk_hid}
\end{figure}

\begin{figure}
    \centering
    \includegraphics[width=5.8cm]{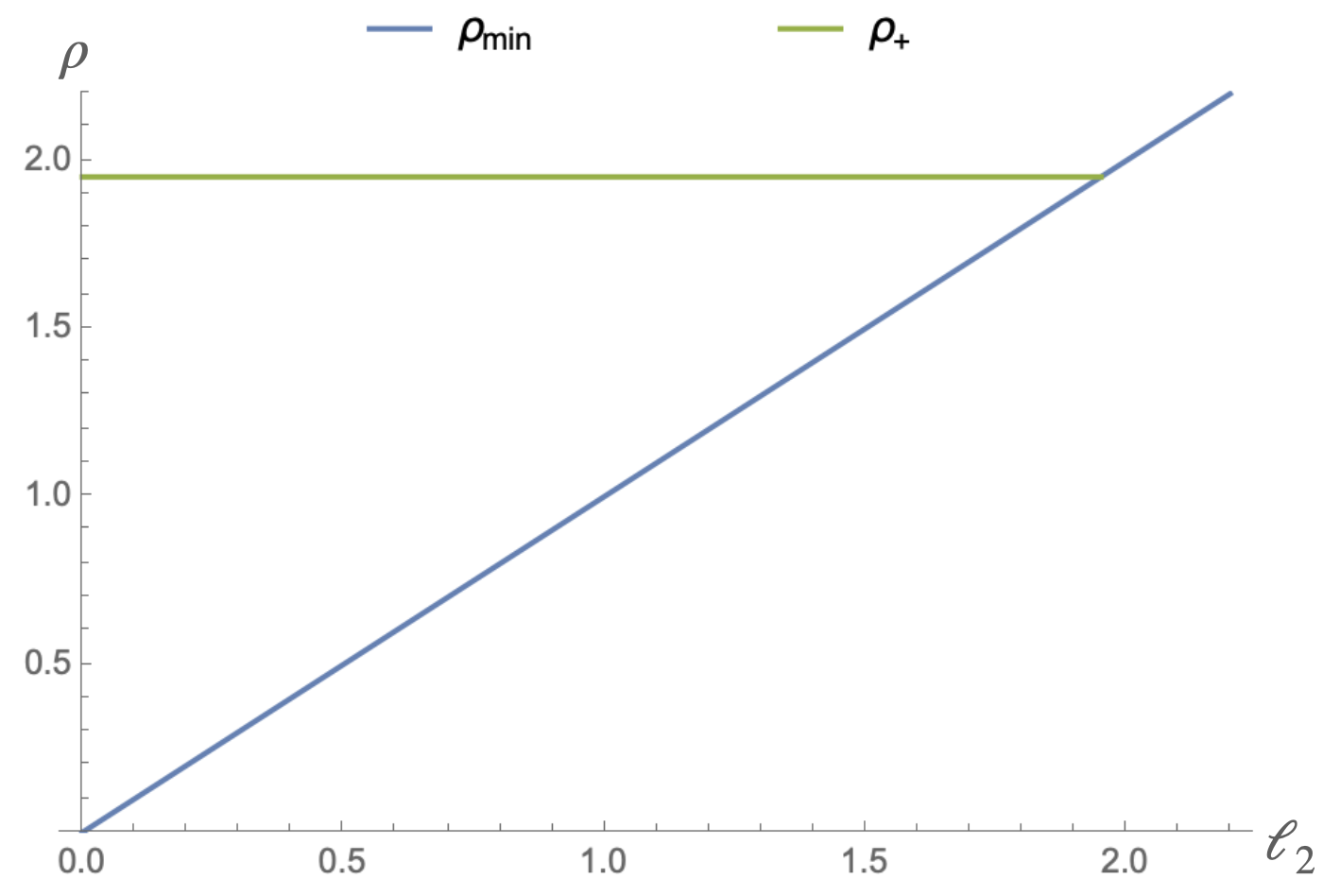}
    \includegraphics[width=5.8cm]{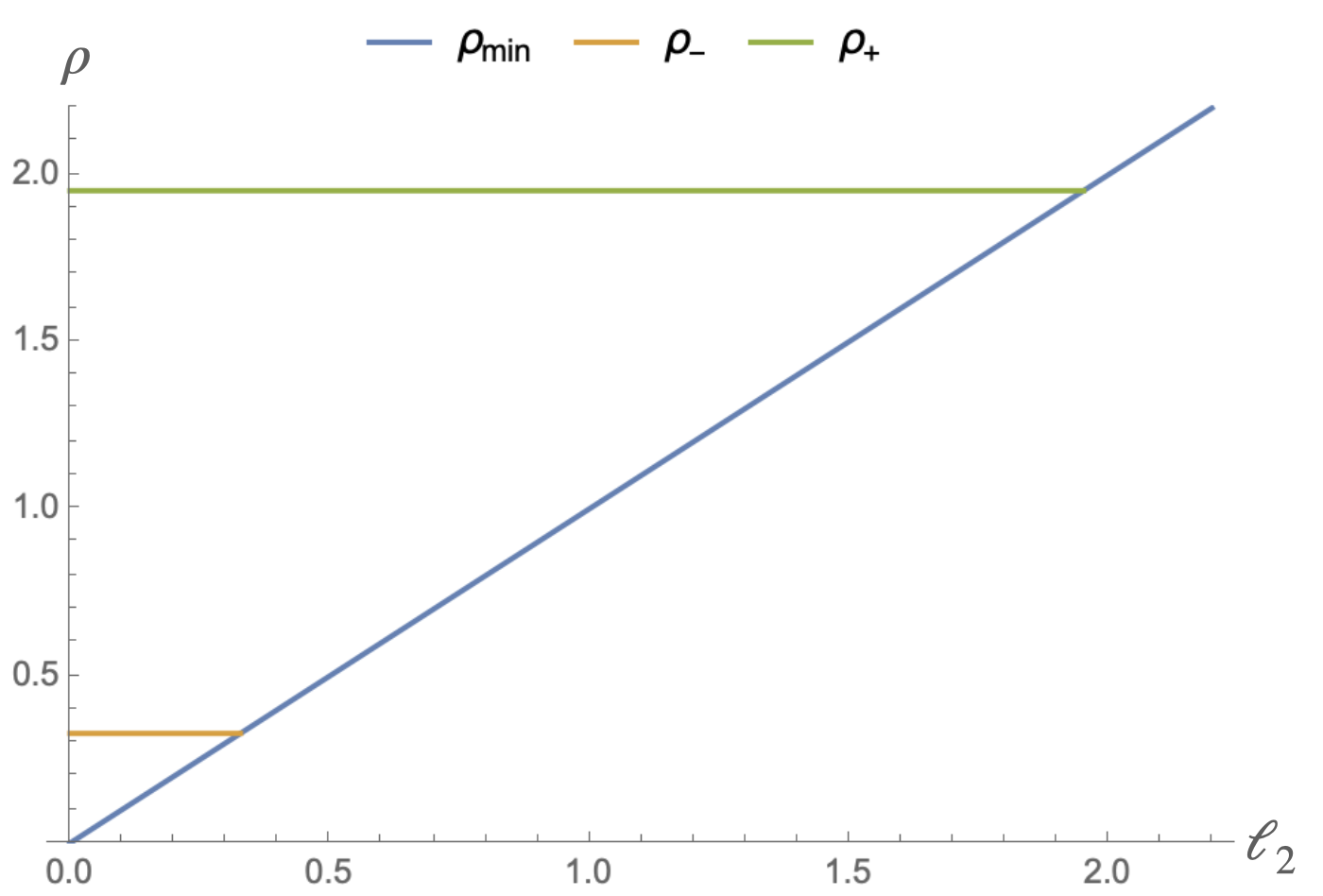}
    \caption{Location of the inner and outer horizons $\rho_\pm$ 
    and the wormhole throat,  
    ($r_{\rm throat}=0$ corresponding to $\rho_{\rm throat}=\ell_2$),
    for $M=1$ and $\ell_1=0$ (left) and $\ell_1=0.3$ (right) 
    as a function of $\ell_2$.}
    \label{F:mixed}
\end{figure}

\begin{itemize}
   
\item
Let us now set $\ell_1<\ell_1^\star \equiv 4 M/(3\sqrt{3})$ and study the situation for different values of $\ell_2$.
We recall that in this situation
\begin{equation}
    r_i(M, \ell_1,\ell_2) = \pm \sqrt{ \rho_i^2(M,\ell_1)-\ell_2^2} . 
\end{equation}
with two distinct roots for $\rho_i(M,\ell_1)$ (remember that we are discarding negative roots).
With reference to Fig.~\ref{F:mixed}, the wormhole throat is always located at $ r=0$, so $\rho_{\rm throat}\equiv\ell_2$. 
For sufficiently small values of $\ell_2$ there will be four horizons located at 
\begin{equation}
    r_i(M, \ell_1,\ell_2) \approx \pm |\rho_i(M,\ell_1)| . 
\end{equation}
This configuration describes, for each side of the universe, a pair of outer and inner horizons shielding a timelike wormhole throat. A Penrose diagram of this spacetime is provided in Fig.~\ref{fig:tmlk_hid}. Increasing the value of $\ell_2$, the throat at $r=0$ moves toward both the inner and outer horizon. 
For $ \ell_2> \min|\rho_i(M,\ell_1)|$ one pair of horizons disappears: there is now only a pair of outer horizons (one for each side of the universe) and the wormhole throat is spacelike, corresponding to a ``black bounce''.  This spacetime is qualitatively equivalent to the hidden wormhole described in Sec.~\ref{SS:hidden}. Finally, for $ \ell_2> \max|\rho_i(M,\ell_1)|$, no horizons are present and the geometry describes a naked wormhole, a globally traversable wormhole. Note that for $\ell_2\to 0$ the standard branches of the Hayward geometry described in Sec.~\ref{sec:Reg-simply1} are recovered.

\item 
Let us now set $\ell_1\to\ell_1^\star \equiv 4M/(3\sqrt{3})$ and study the situation for different values of $\ell_2$. 
There are now two value of $r$ corresponding to the degenerate positive root of $\rho$ 
\begin{equation}
    r_E = \pm \sqrt{ \left(4M\over3\right)^2 - \ell_2^2} ,
\end{equation}
As long as $\ell_2$ is sufficiently small ($\ell_2< 4M/3$) these will be real, corresponding to a pair of extremal horizons (one for each side of the throat). For $\ell_2 > 4M/3$ one has a horizonless object with a timelike throat at $r=0$. For $\ell_2\to0$ one recovers the simply connected extremal geometries discussed in Sec.~\ref{sec:Reg-simply2}.

\item
For $\ell > \ell_1^\star\equiv 4 M/(3\sqrt{3})$ there is only one real root for $\rho$ which being negative should be discarded. Hence, no horizons are present and one recovers a naked traversable wormhole. In the limit $\ell_2\to 0$ one recovers as expected a horizonless simply connected geometry as described in Sec.~\ref{sec:Reg-simply3}.
\end{itemize}

 %=================================
%==================================

%================================================
\section{Dynamical geometries}
%================================================

The static geometries described above are non-physical, in the sense that these cannot describe situations in which black holes are formed in a gravitational collapse process. The static nature of these solutions causes further problems, such as the existence of Cauchy horizons, that are generally not present in time-dependent situations. Hence, it is important to keep in mind that we will use these static geometries carefully, as useful building blocks of time-dependent situations, but do not take seriously some of the issues that appear only in the (unphysical) static limit.

With this in mind, the geometries we will be working with in this section can be written in Eddington-Finkelstein form as
\begin{equation}
    ds^2=-F(v,r)\,dv^2+2\,dv\,dr+\rho^2(v,r)\,d\Omega^2\,,
\end{equation}
where $F(v,r)$ and $\rho(v,r)$ are obtained from the expressions in Eq.~\eqref{eq:example} when implicit $v$-dependence through the functions $M(v)$, $\ell_1(v)$ and $\ell_2(v)$ is allowed: 
\begin{equation}
    F(v,r)=1-\frac{2M(v)\rho^2(r)}{\rho^3(r)+2M\ell_1^2(v)}\,,\qquad\qquad \rho^2(v,r)=r^2+\ell_2^2(v).
\end{equation}
The three functions $M(v)$, $\ell_1(v)$ and $\ell_2(v)$ play three very different roles:
\begin{itemize}
\item $M(v)$ is the Bondi mass, and thus determines the total amount of mass contained in the corresponding spacetime at a given moment $v$. It has a pronounced effect on the location of the outer horizon, if existing.
\item $\ell_1(v)$ has a pronounced effect on the location of the inner horizon, if existing.
\item $\ell_2(v)$ has a pronounced effect on the location of the wormhole throat.
\end{itemize}
Hence, these geometries can describe the formation of outer/inner horizons and wormhole throats that evolve dynamically. Let us discuss the possible dynamical behaviors that can arise, starting from the topological classification that was also useful in static situations.

%==========================================================
\subsection{Dynamical regularization in simply connected topologies}
%==========================================================

Within the family of dynamical geometries we are considering as an example, the simply connected topologies are given by $\ell_2=0$, so that we have the metric
\begin{equation}
    F(v,r)=1-\frac{2M(v)\rho^2(r)}{\rho^3(r)+2M\ell_1^2(v)}\,,\qquad\qquad \rho^2(r)=r^2.
\end{equation}

This general metric contains metrics previously analyzed in the literature.
In particular, if we drop the time dependence in $\ell_1(v)$, writing it as a constant quantity, we recover the metric used by Hayward in his analysis of evaporating regular black holes~\cite{Hayward:2005gi}. In fact, it is interesting to note that the assumption of varying $M(v)$ with $\ell_1\neq 0$ held fixed has been routinely used in the analysis of the dynamics of regular black holes (see e.g.~\cite{Frolov:2014jva,Frolov:2016gwl,Frolov:2017rjz,Hayward:2005gi,Simpson:2018tsi,Simpson:2019cer,Simpson:2019mud}). 
%\blue{[There should be more references we can add!]}

That the possible time evolution of $\ell_1$ has been ignored is likely both due to an implicit assumption that this quantity must remain constant, as well as technical limitations in the analysis of semiclassical backreaction:

\begin{itemize}
\item
The assumption that $\ell_1$ must remain constant comes from its interpretation as a fundamental scale set by quantum gravity (e.g. the Planck scale)~\cite{Garay:1994en, Booth:2018xvb}. However, this quantity is just a dynamical scalar contained in the metric and, therefore, while it may be reasonable that its initial value can be fixed according to these considerations, there is no strong reason to discard a subsequent dynamical evolution towards different values.
\item The analysis of the backreaction of quantum fields on black holes have so far typically been limited to the region around and outside outer horizons~\cite{Birrell:1982ix,Frolov:1998wf}, with relatively little work on investigating the interior, although some analyses using toy models~\cite{Mersini-Houghton:2014zka,Mersini-Houghton:2014cta}, as well as the actual renormalized stress-energy tensor~\cite{Abedi:2015yga,Barcelo:2020mjw,Barcelo:2022gii} have been carried out. Focusing on the region around and outside the outer horizon leads to the standard result of evaporation due to the emission of Hawking radiation~\cite{Hawking:1975vcx}, which is described as the time dependence in $M(v)$. The analysis of backreaction inside the black hole, and in particular around inner horizons, presents some technical limitations, though recent attempts at analyzing these issues have shown~\cite{Barcelo:2020mjw,Barcelo:2022gii} that backreaction turns into a time dependence for $\ell_1(v)$. 
\end{itemize}

In summary, given our current knowledge, it is reasonable to expect that semiclassical backreaction on regular black holes will lead to time dependence in both scales $M(v)$ and $\ell_1(v)$. The former scale controls the dynamical evolution of the geometry around the outer horizon, while the latter scale controls the dynamical evolution of the geometry around the inner horizon. The precise dynamical evolution of regular black holes cannot be determined without a better understanding of the underlying dynamical laws. 

\begin{figure}
    \centering
    \includegraphics[height=7cm]{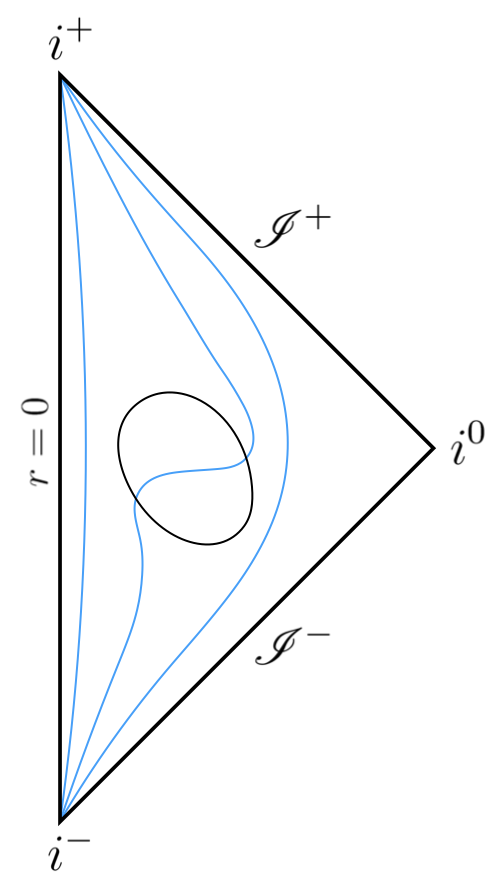}  
    \caption{Dynamical regular black hole formed by gravitational collapse. Here we allow $M(v)$ and $\ell_1(v)$ to be time dependent, but keep $\ell_2=0$. Once formed, the trapped region disappears in finite time, albeit the latter is undetermined (missing a specific dynamics) and hence can be also very large, formally even infinite. The blue lines correspond to lines of constant radius and are timelike outside the trapped region and spacelike inside it. Contrary to the eternal case, this spacetime is globally hyperbolic.}
    \label{fig:dyn_rbh}
\end{figure}

Regardless of this uncertainty, we do know that the resulting geometry must belong to the classes \{\ref{class1}, \ref{class3}\}. We provide below, in Fig.~\ref{fig:dyn_rbh}, a Penrose diagram for the physically more relevant class \ref{class1}. The diagram for class \ref{class3} can be seen as the everlasting limit of this one. Hence, while not all details are fixed (such as the duration of the horizon structure in class \ref{class1}), the qualitative behavior associated with this regularization mechanism is well understood. Among qualitative universal features, stands out the lack of Cauchy horizons. The existence of Cauchy horizons are associated with the unphysical restriction to static geometries, but these are replaced by dynamical inner horizons in dynamical situations.

%%%%%%%%%%%%%%%%%%%%%%%%%%%
%%%%%%%%%%%%%%%%%%%%%%%%%%%%%
\bigskip
%\blue{\bf *** Definitely need a Penrose diagram here... ***}

%==========================================================
\subsection{Dynamical regularization in non-simply connected topologies}
%==========================================================

Within the family of geometries we are considering as an example, the non-simply connected topologies are given by $\ell_1=0$, so that we have the metric
\begin{equation}
    F(v,r)=1-\frac{2M(v)}{\rho(v,r)}\,,\qquad\qquad \rho^2(v,r)=r^2+\ell_2^2(v).
\end{equation}

These geometries can describe the creation of a wormhole at a finite time, by including either non-smooth functions or non-analytic smooth functions in the definition of $\ell_2(v)$. As mentioned above, this leads to the breakdown of global hyperbolicity.
(The special case of varying $M(v)$ with $\ell_2\neq 0$ held fixed is discussed extensively in Ref.~\cite{Simpson:2019cer}.)

In principle, the lack of global hyperbolicity makes these geometries less appealing. However, the creation of wormholes (and thus, the associated violation) can happen at arbitrarily small scales, perhaps associated with quantum gravity. In a similar way as we discussed with the function $\ell_1(v)$ for regular black holes, in the present case $\ell_2(v)$ can start taking Planckian values and then flowing dynamically to macroscopic values. 

Whether or not this behavior is compatible with known laws of physics (e.g. semiclassical physics) is still unknown. However, if any geometries of this kind is realized, it must belong to the classes \{\ref{class2}, \ref{class4}\}. 
Penrose diagrams for this class of spacetimes are generically multi-sheeted and hence difficult to draw (this is a straightforward consequence of the topology-change associated to these geometries). For this reason we do not report them here.

%==========================================================
\subsection{Dynamical regularization in ``mixed'' cases}
%==========================================================

One could let all three of the parameters, $M(v)$, $\ell_1(v)$, and $\ell_2(v)$,  become time dependent. This would yield a superset of all cases considered above. 
While technically somewhat more complex, no really new issues of principle are involved. All non-trivial ($\ell_2(v)\neq0$) mixed cases are not globally hyperbolic. Hence, as with the models of dynamical regularization in non-simply connected topologies, topology change must occur in gravitational collapse for these spacetimes to provide a viable description.

%==============================================
\section{Discussion and wrap-up}
%==============================================

In wrapping up this chapter let us reflect on the major points we have considered.
Firstly, a purely kinematic ``geometrographic'' analysis --- inspired by the desire to avoid singularities --- already places significant constraints on just how singularity avoidance might be achieved. 
To merely appeal to a generic ``quantum smoothing'' and 
hiding all of the details into the ``too hard basket'' does not make a good contribution to knowledge;
there is a reasonably well fleshed out taxonomy of kinematically acceptable scenarios that can plausibly be justified from first principles. 

We have found good reason to focus on spacetime geometries of the form
\begin{equation}
    ds^2=-F(v,r)\,dv^2+2\,dv\,dr+\rho^2(v,r)\,d\Omega^2\,.
\end{equation}
Here $F(v,r)$ and $\rho(v,r)$ are required to satisfy certain conditions (discussed above) to keep the spacetime regular. A sufficiently broad class of models is obtained by letting the 2-variable functions $F(v,r)$ and $\rho(v,r)$ depend implicitly on $v$ 
 through the three 1-parameter functions $M(v)$, $\ell_1(v)$ and $\ell_2(v)$ by setting: 
\begin{equation}
    F(v,r)=1-\frac{2M(v)\rho^2(r)}{\rho^3(r)+2M\ell_1(v)^2}\,,\qquad\qquad \rho^2(v,r)=r^2+\ell_2^2(v).
\end{equation}
This explicit class of models is still good enough to cover all of the general classes of spacetimes determined to be of interest from our ``geometrographic'' arguments.

Focusing on these specific models allows us to say quite a bit about candidate regular black holes --- and their extremal limits and horizonless counterparts. In the longer run, such considerations are also of  use for further phenomenological analyses and for planning future 
observational projects~\cite{ Barack:2003fp, Barausse:2020rsu, Berti:2005ys, Cardoso:2019rvt}.

\begin{acknowledgement}
RCR acknowledges financial support through a research grant (29405) from VILLUM fonden.
FDF acknowledges financial support by Japan Society for the Promotion of Science Grants-in-Aid for international research fellow No. 21P21318. 
SL acknowledges funding from the Italian Ministry of Education and  Scientific Research (MIUR)  under the grant  PRIN MIUR 2017-MB8AEZ. 
CP acknowledges the financial support provided under the European Union's H2020 ERC, Starting Grant agreement no.~DarkGRA--757480 and support under the MIUR PRIN and FARE programmes (GW- NEXT, CUP: B84I20000100001).
MV was supported by the Marsden Fund, via a grant administered by the Royal Society of New Zealand. 
%If you want to include acknowledgments of assistance and the like at the end of an individual chapter please use the \verb|acknowledgement| environment -- it will automatically render Springer's preferred layout.
\end{acknowledgement}

%%%%%%%%%%%%%%%%%%%%%%%%%%%%%%%%%%%%%%%%%%%%%%%%%%%%%%%%%

%================================================
\section*{Appendix 1: Extremal horizons}
%================================================

A general feature implicit in the discussion above is that special things happen to the spacetime geometry at horizons, 
and that even more special things happen at extremal horizons. 
However different special things might happen for inner \textit{versus} outer horizons. 
In this appendix we shall seek to present a coherent overview of this topic.
(In a somewhat similar vein, it has been known for some time that special things happen at wormhole throats~\cite{Hochberg:1997wp}.) 

We find it convenient to work with static spacetimes in area coordinates:
\begin{equation}\label{Eq1}
    ds^2 = -e^{-2\Phi(r)}\left(1-\frac{2m(r)}{r}\right)dt^2 + \frac{dr^2}{1-\frac{2m(r)}{r}} + r^2\,d\Omega^2_{2}.
\end{equation}
The horizons are located at solutions (if any) of the equation $r_H = 2 m(r_H)$. If we are dealing with a wormhole throat, we will need two coordinate systems of this type, carefully matched at the throat~\cite{Visser:1995cc}.

A purely geometrical result is that in a suitable orthonormal basis~\cite{Visser:1995cc}
\begin{equation}
G_{\hat t \hat t} = {2m'(r)\over r^2}; \qquad\qquad
%\end{equation}
%\begin{equation}
G_{\hat r \hat r} = - {2m'(r)\over r^2}+ \left(1-{2m(r)\over r} \right){2\Phi'(r)\over r};
\end{equation}
Thence at any horizon (inner or outer, extremal or non-extremal) one has
\begin{equation}
G_{\hat t \hat t} +  G_{\hat r \hat r}  \quad\to\quad 0.
\end{equation}
This on-horizon ``enhanced symmetry'' for the Einstein (and Ricci) tensors is a recurring theme in near horizon physics~\cite{Martin-Moruno:2021niw}. 
Another useful and very general result is that the surface gravity is~\cite{Visser:1992qh}:
 \begin{equation}
\kappa_H  = e^{-\Phi(r_H)} \; {1-2m'(r_H)  \over 2r_H }.
\end{equation}
An extremal horizon is characterized by the vanishing of the surface gravity which we see requires $2m'(r_H)=1$. 

At any extremal horizon (outer or inner) the Einstein tensor (and Ricci tensor) become particularly simple. Specifically
\begin{equation}
G_{\hat t \hat t|r_E} = - G_{\hat r \hat r|r_E} = {1\over r_E^2};
\qquad\qquad
G_{\hat\theta\hat\theta|r_E} = G_{\hat\phi\hat\phi|r_E} = -{m''(r_E) \over r_E}.
\end{equation}
In fact at any extremal horizon all orthonormal components of the Riemann tensor are proportional to either $1/r_E^2$, or $m''(r_E) /r_E$, or are zero. Thence at any extremal horizon all of the polynomial curvature invariants are simply multi-nomial functions $f(1/r_E^2,m''(r_E) /r_E)$ of these two quantities.
Furthermore at any extremal horizon all nonzero orthonormal components of the Weyl tensor are simply miltiples the single quantity  $(1+r_E m''(r_E))/r_E^2$.
So the spacetime geometry simplifies quite drastically on any extremal horizon (either outer or inner).

Finally, what can we say about $m''(r_E)$? This will depend on how many non-extremal horizons merge to yield the extremal horizon of interest. If two non-extremal horizons merge then $m''(r_E)\neq0$; if three (or more) non-extremal horizons merge then $m''(r_E)=0$. 
For instance in the extremal Reissner-Nordstr\"om geometry (where two horizons merge) 
we have $m(r) = m - {1\over2}q^2/r$ and at extremality we obtain $m''(r_E)= -q^2/r_E^3 = -1/r_E < 0$. 
In contrast for the extremal inner horizons explored in Ref. \cite{Carballo-Rubio:2022kad} we have three merging horizons, $m(r)$ is a rational quartic and it is easy to check that $m''(r_E)=0$. 

In short, although it is perhaps not all that well appreciated, geometrically it is guaranteed that very special  things happen at all extremal horizons; and these special properties will have a role to play in both phenomenology and in stability analyses for RBHs.

%================================================
\section*{Appendix 2: Light rings}
%================================================

From the discussion above we have seen that interesting  things happen for light rings in extremal, near-extremal, and super-extremal objects. 
See also~\cite{Bargueno:2022vkf, Hod:2022mys, Zhong:2022jke}.
That something unusual happens with light rings in the extremal limit can already be deduced from the very simple and explicit example of Reissner--Nordstr\"om spacetime. Since this situation already captures the key features of the discussion with an absolute minimum of fuss, we present some brief pedagogical comments here, before looking at the general situation.

%=================================================
\subsection*{Reissner--Nordstr\"om light-rings}
%=================================================

The  Reissner--Nordstr\"om spacetime in area coordinates is
\begin{equation}
d s^2 = -(1-2m/r+q^2/r^2) d t^2 +{d r^2\over 1-2m/r+q^2/r^2} + r^2d \Omega^2
\end{equation}
It is a quite standard result that in Reissner--Nordstr\"om spacetime the light rings can be found by inspecting the effective potential
\begin{equation}
V(r) = \left(1-{2m\over r}+{q^2\over r^2} \right) \left({L^2\over r^2}\right)
\end{equation}
The circular photon orbits are located at $r_c$ such that $V'(r_c)=0$, and stability depends on the sign of $V''(r_c)$. 
If $V''(r_c)>0$ then the light ring is stable; if $V''(r_c)<0$ then the light ring is stable;
if $V''(r_c)=0$ then the light ring exhibits neutral (marginal) stability.

Unfortunately, the potential $V(r)$ \textit{is not unique}, a circumstance which can sometimes cause confusion.
Indeed, let $F(x)$ be any monotone increasing function and define $\tilde V(r) = F(V(r))$. 
Then 
\begin{equation}
\tilde V(r)' = F'(V(r)) \;V'(r); \qquad \tilde V(r)'' = F''(V(r))\; [V'(r)]^2 + F'(V(r)) \;V''(r).
\end{equation}
So the extrema $r_c$ of $V(r)$ coincide with extrema of $\tilde V(r)$. 
Furthermore, at these extrema one has $sign\{\tilde V''(r_c)\} = sign\{V''(r_c)\}$. 

To locate the light rings we note
\begin{equation}
V'(r) = {2L^2\over r^5}(3mr-2q^2-r^2),
\end{equation}
and
\begin{equation}
V''(r) = {2L^2\over r^6}( 3r^2+10q^2-12mr).
\end{equation}
The  outer and inner horizons are located at
\begin{equation}
r_H = m \pm \sqrt{m^2-q^2}.
\end{equation}
The   outer and inner  light rings are located at
\begin{equation}
r_c = {3m\over2} \pm {\sqrt{9m^2-8q^2}\over 2}.
\end{equation}
Distinct inner and outer light rings exist for $9m^2 > 8q^2$, and merge at $9m^2 = 8q^2$. that is, beyond extremality.  The light rings merge at $r_c= {3m\over2}$. 

\begin{figure}[!htb]
\begin{center}
\includegraphics[scale=0.4]{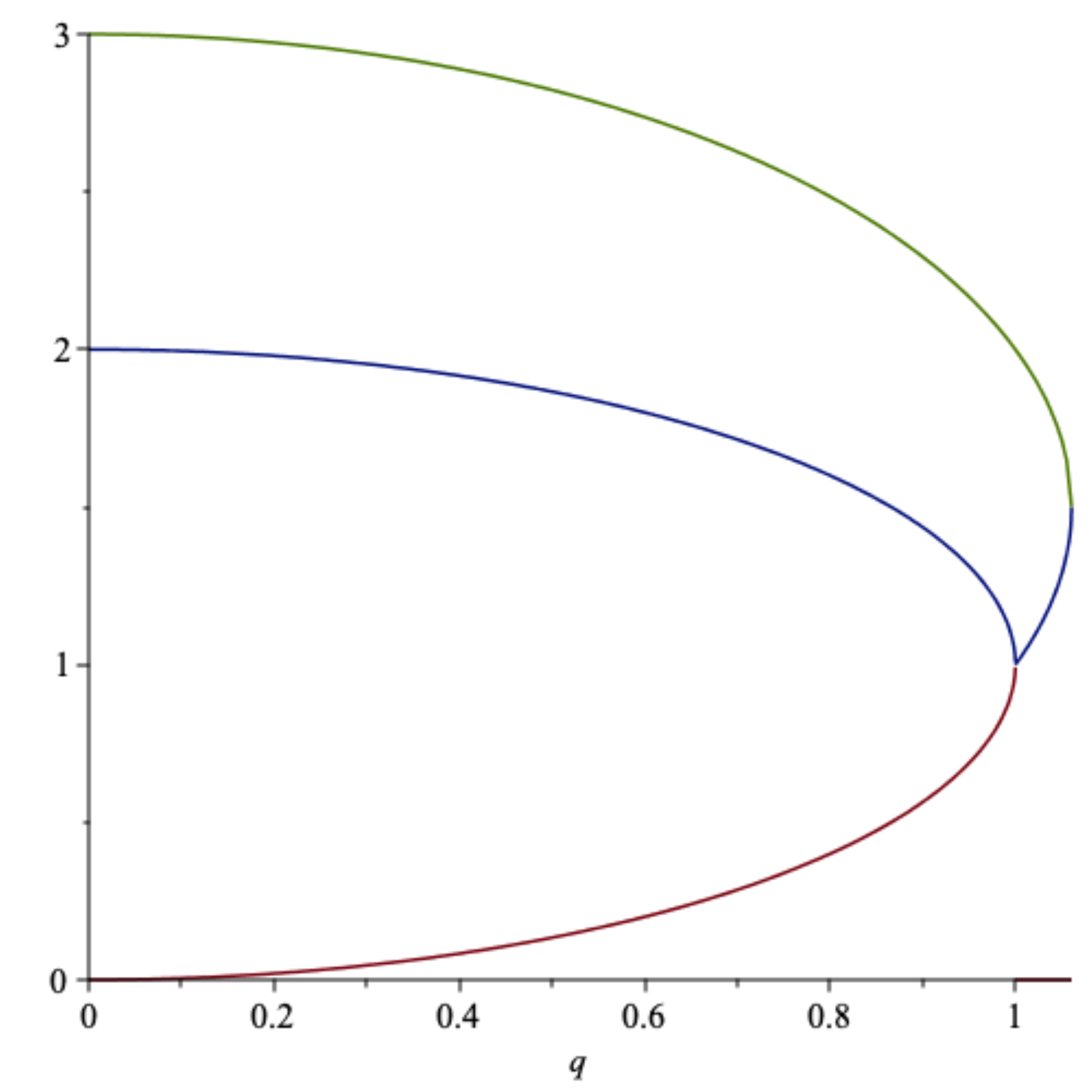}
\caption{Reissner-Nordstr\"om inner and outer horizons, and inner and outer light rings.}
\label{F:RN}
\end{center}
\end{figure}

\begin{itemize}
\item
At the outer light ring
\begin{equation}
V''(r_c) = - {64 L^2 \sqrt{9 m^2-8 q^2} \over (3m + \sqrt{9 m^2-8 q^2})^5 } < 0,
\end{equation}
so the outer light ring is always unstable.
\item
At the inner light ring
\begin{equation}
V''(r_c) =  {64 L^2 \sqrt{9 m^2-8 q^2} \over (3m - \sqrt{9 m^2-8 q^2})^5 } > 0,
\end{equation}
so the inner light ring is always stable.
\item
At $r_c= {3m\over2}$, where the light rings merge, $9 m^2=8q^2$ so $V''(r_c) =  0$,
and the merged light ring exhibits neutral stability.
\end{itemize}
At extremality ($m=|q|$) the light rings are formally located at 
\begin{equation}
r_c= {3m\over2} \pm {m\over2} = \{ m, 2m\}.
\end{equation}
Here $r_c=2m$ corresponds to a true light ring, while $r_c=m$ represents the light sheet defining the extremal horizon. (There is now no angular motion, so this is not a ``ring''.) 
At extremality ($m=|q|$) for the outer light ring
\begin{equation}
r_c = 2m = 2 r_H; \qquad V''(r_c) \to  -{ L^2 \over 16 m^4 } < 0.
\end{equation}
At extremality ($m=|q|$) for the inner light sheet
\begin{equation}
r_c=m= r_H; \qquad V''(r_c) \to  +{ 2L^2 \over m^4 } > 0.
\end{equation}
So the extremal horizon is a stable light sheet.
Perhaps counter-intuitively, the fact that the light sheet is stable will destabilize the spacetime --- since the light sheet is stable, massless particles cam pile up there; eventually back-reaction will become large, and the spacetime detabilizes. 
The situation is summarized in Fig.~\ref{F:RN}. 

The key observation here is that the Reissner-Nordstr\"om spacetime is already subtle enough to exhibit a stable light ring at extremality, and multiple light rings in a small region beyond extremality.  
This does have implications for more general RBHs, since one can always cut off the core of the 
Reissner-Nordstr\"om spacetime at some $r_{core} < |q|$ and replace it with a Reissner-Nordstr\"om-inspired RBH that would then (by construction) exhibit exactly the same light rings as the Reissner-Nordstr\"om spacetime  itself. 
In short, the existence of unstable light rings exterior to generic extremal black holes should not really come as a surprise. 

%======================
%======================

%=================================================
\subsection*{Generic light-rings}
%=================================================

What can we say about light rings in the generic case?
We already have rather good intuition based on what we saw happening for 
Reissner-Nordstr\"om spacetime. 

%\enlargethispage{60pt}
For any geometry of the form given in Eq.~(\ref{Eq1}) it is easy to check that the effective potential 
governing the light rings is
\begin{equation}
V(r) = e^{-2\Phi(r)} \left(1-{2m(r)\over r}\right) {L^2\over r^2}.
\end{equation} 
It is then easy to check that
\begin{equation}
V'(r) = e^{-2\Phi(r)}\, {2L^2\over r^4}\, \left( \{3m(r)-r m'(r)-r\}- \Phi'(r) r^2(1-2m(r)/r) \right).
\end{equation}
Purely geometrically this leads to
\begin{equation}
\label{E:4upper}
V'(r) = e^{-2\Phi(r)}\, {2L^2\over r^4}\,  \{3m(r) -r  + r^3 G_{\hat r\hat r}(r) \}.
\end{equation}
This can also be written as
\begin{equation}
V'(r) = e^{-2\Phi(r)}\, {2L^2\over r^4}\,  \{3m(r) -r -rm'(r) + r^3 [G_{\hat t\hat t}(r) + G_{\hat r\hat r}(r)] \}.
\end{equation}
Furthermore one can easily verify that
\begin{equation}
V''(r) =  e^{-2\Phi(r)} {2L^2\over r^5}\{3r-12m(r)+6 r m'(r) -r^2 m''(r)\} -2\Phi'(r) V'(r) -2\Phi''(r)  V(r). 
\end{equation}
Thence at any light ring that might be present the condition $V'(r_c)=0$ implies
\begin{equation}
r_c = {3m(r_c) + r_c^3 [G_{\hat t\hat t}(r_c) + G_{\hat r\hat r}(r_c)] \over 1 - m'(r_c) }
\end{equation}
We now wish to self-consistently bound the location of possible solutions to this equation to determine whether a light ring exists for $r_c > r_H$.
The purely geometrical \textit{null convergence condition} (guaranteeing the convergence of null geodesics), when applied to the radial null geodesics, would imply 
\begin{equation}
[G_{\hat t\hat t}(r) + G_{\hat r\hat r}(r)] \geq 0.
\end{equation}
If the Misner-Sharp quasi-local mass is non-decreasing outside the horizon then this would imply $m'(r) \geq 0$. Combining, if we assume the light ring exists, then its location is bounded below by:
\begin{equation}
r_c \geq 3m(r_c) \geq 3 m(r_H) = {3\over2}\; r_H.
\end{equation}
To establish an upper bound we start from Eq. (\ref{E:4upper}). Setting $V'(r_c)\to0$ yields
\begin{equation}
r_c = 3m(r_c) + r_c^3 G_{\hat r\hat r}(r_c) .
\end{equation}
Then from the DCC (dominant convergence condition): $|G_{\hat r\hat r}| \leq |G_{\hat t\hat t}| = {2m'/r^2}$ we see
\begin{equation}
r_c \leq  3m(r_c) + 2 r_c m'(r_c).
\end{equation}
We need one more condition to get a useful bound: $(m(r)/r^3)' < 0$, implying $m'(r) < 3 m(r)/r$. (This condition corresponds to the volume-averaged density decreasing as one moves outwards, and is a very popular condition used in building relativistic and Newtonian models.) Then
\begin{equation}
r_c \leq  9m(r_c) \leq 9 m_\infty.
\end{equation}
Overall, under plausible structural conditions, and assuming existence of the light ring, we have
\begin{equation}
{3\over2} r_H \leq r_c \leq 9 m_\infty.
\end{equation}
However, proving actual existence of the light rings is slightly more subtle, and requires slightly different arguments for outer-non-extremal and outer-extremal horizons.

%=================================================
\subsubsection*{Outer non-extremal horizons}
%=================================================

For an outer non-extremal horizon in terms of the surface gravity we can calculate
\begin{equation}
V'(r_H) = {2 L^2 e^{-\Phi(r_H)} \; \kappa_H\over r_H^2} >0.
\end{equation}
On the other hand for an asymptotically flat geometry, at large $r$ we will have $m(r)=m_\infty+O(1/r)$ 
and $\Phi(r)=O(1/r)$ whence asymptotically\footnote{There are additional complications if one abandons asymptotic flatness. For instance in asymptotically de Sitter spacetimes one also encounters OSCOs, outermost stable circular orbits~\cite{Berry:2020ntz, Boonserm:2019nqq}.} 
\begin{equation}
V'(r) = -{2 L^2 \over r^3} + O\left(1\over r^4\right) < 0.
\end{equation}
The sign flip guarantees that there will be at least one light ring somewhere between the outer horizon and spatial infinity.

%=================================================
\subsubsection*{Outer extremal horizons}
%=================================================

For any extremal horizon  we can calculate
\begin{equation}
V'(r_H) = 0; \qquad V''(r_H) = -{2L^2 e^{-2\Phi(r_H)} m''(r_H)\over r_H^3}.
\end{equation}
If this is to be an outer extremal horizon then we must have $ m''(r_H)<0$ and so $V''(r_H)>0$.
But then $V'(r)>0$ in the region immediately above the horizon. On the other hand, for any asymptotically flat geometry we still have
\begin{equation}
V'(r) = -{2 L^2 \over r^3} + O\left(1\over r^4\right) < 0.
\end{equation}
The sign flip again guarantees that there will be at least one light ring somewhere between the outer horizon and spatial infinity.

%=================================================
\subsubsection*{Regular horizonless objects}
%=================================================

For a regular horizonless object, (\textit{cf.} a super-extremal Reissner-Nordstr\'om geometry with a regularized core at $r_{\rm core}< |q|$), at short distances we would demand $m(r)=O(r^3)$ and $\Phi(r) = O(r^2)$. Consequently
\begin{equation}
V'(r) = -{2 L^2 \over r^3} + O\left(1\right) < 0,
\end{equation}
while at large distances
\begin{equation}
V'(r) = -{2 L^2 \over r^3} + O\left(1\over r^4\right) < 0.
\end{equation}
There is now no sign flip and consequently there must be an even number (possibly zero) of light rings.

\end{document}